\documentclass[aps,prl,twocolumn,epsfig,showpacs]{revtex4}

\usepackage[centertags]{amsmath}
\usepackage{amssymb}
\usepackage{amsthm}
\usepackage{graphicx,subfigure}
\usepackage{epsfig}

\begin{document}

\title{Stochastic dynamics of two-step processes with harmonic potential}

\author{Jyotipriya Roy$^1$, Chitrak Bhadra$^1$, Debapriya Das$^1$, Dhruba Banerjee$^1$ and Deb Shankar Ray$^2$}

\affiliation{
$^1$ Department of Physics, Jadavpur University, Kolkata, India. \\
$^2$ Department of Physical Chemistry, \\
Indian Association for the Cultivation of Science, Kolkata, India.}

\begin{abstract}
\noindent
In this paper we address the one-dimensional problem of stochastic renewal in different damping environments. An ensemble of particles with some specified initial distribution in phase space are allowed to evolve stochastically till a certain instant of time (say $\tau$), when a restoring force is applied to bring them back to some point in configuration space. The physical quantities of interest that have been studied are the survival probability and the first passage distribution for return to the specified target point. We observe nontrivial dependence of these quantities on $\tau$ as well as on the width of the initial distribution, which has been taken to be Gaussian in position and velocity.
\end{abstract}

\maketitle

\section{INTRODUCTION}

\noindent Ever since Einstein \cite{ein} initiated the study of stochastic processes through his seminal paper on Brownian motion, studies relating to the effects of randomness in physical systems have evolved astronomically \cite{chand,ornuhl,hang,wax}. A basic paradigm for studying stochastic processes involves the specification of some initial distribution (or concentration) of an ensemble of particles, and then studying how this distribution evolves according to some dynamical equation, appropriate for the situation under consideration. Even if one is not concerned with quantum effects, wide variations in situations are encountered on taking into account the different degrees of damping offered by the environment \cite{kram,dsr}. The importance of the fluctuation-dissipation relation that addresses the interplay of damping and random forcing can hardly be overemphasized in this context \cite{red,risk,gard}. This paper addresses, in the afore-mentioned lines, the dynamics of stochastic processes that involve two distinct but inter-connected stages, distinguished by the presence or absence of some confining potential.

\noindent We pose the problem in this way: a particle of mass $m$ starts evolving stochastically from a starting point specified by some initial probability distribution. In this process it behaves like a free Brownian particle moving through a thermal bath with damping and random forcing complementing each other. After a certain instant of time, called the resetting time (prescribed by the observer), a harmonic trap is suddenly set up, so that, under the action of the harmonic force, the particle is dragged back to some target point (say, the origin). Therefore, the entire journey of the particle is temporally divided into two distinct phases, one in which the motion is purely stochastic (through a thermal bath with either low or heavy damping), and the other where the motion is dictated by the potential. There can be three possible situations (in this second phase of the motion) by which the particle can come back: first, the particle is brought back deterministically by the potential, i.e., the stochastic bath is not involved in this part of the process and hence the governing equation for probability evolution is a pure Liouville equation with random initial conditions; second, the particle is brought back in a weak-damping environment, and hence in this case both the phases of the motion are governed by Kramers type equations; and third, the resetting is done in a heavy-damping environment, in which the inertial term in the corresponding Langevin equation is dropped (thus leading to Smoluchowski equation) and the probability evolves according to a Fokker-Planck equation in configuration space only. We take up each of these three cases (though not in the aforementioned order) one by one, in the following three sections.

\noindent In recent years, stochastic resetting of random searchers have attracted the interest of some respected groups of statistical physics \cite{satya1,satya2}. This field, whose mathematical foundations lie in renewal theory of probability, are useful in studying a variety of physical situations in physics, chemistry and other natural sciences \cite{fell,van}. These problems have mostly focussed on finding the optimal returning or resetting rates of random searchers that have set out from some point in real space, to search something, until they are reset to their original starting point, or some other point. This arena of stochastic processes is rich in its own right as it connects and addresses many real world situations, and hence, comprehensive studies including extensions to higher dimensions have also been done \cite{satya3}. Our interest in this paper is not to address the subject of stochastic resetting directly. However, a close examination of what has been just said reveals, that this resetting process actually involves two consecutive steps: one, the searcher goes out to search, and two, the searcher is drawn back to the starting point. Usually, the dynamics of the second phase does not find much importance or relevance in the afore-mentioned studies on stochastic resetting. In this work, we make an equation-of-motion approach to such two-step processes in different damping environments, thus incorporating a definite force which actually brings the searcher back to its starting point during what we call the second phase of the dynamics.

\noindent
A small note on nomenclature: When we name a process as a ``Kramers-Liouville process'', we mean that the first phase of the motion is governed by Kramers equation while the second phase is governed by Liouville equation. On the other hand, a ``pure Kramers process'' would imply that both the phases of the motion are governed by Kramers-type equations, albeit of different structures. Similarly, a ``pure Smoluchowski process'' means that both the phases are governed by Smoluchowski type equations.

\renewcommand{\theequation}{1.\arabic{equation}}
\setcounter{equation}{0}
\section{I. The Heavy Damping Case: a pure Smoluchowski process}

\noindent
We begin with the simplest of the three cases, viz., the two-step process taking place in a heavy damping environment. We start out with an initial Gaussian distribution of the particles in configuration space, given by

\begin{equation}
P_1 (x,t=0) = \frac{1}{\sigma \sqrt{2\pi}} \exp \left( -\frac{x^2}{2\sigma^2} \right)
\label{eq1.1}
\end{equation}

\noindent where, $\sigma$ represents the half-width of the initial Gaussian distribution. The subscript $1$ of $P_1$ symbolizes that we are in the $1$st phase of the motion, a pure diffusive phase governed by the equation

\begin{eqnarray}
\frac{\partial P_1}{\partial t} &=& G \frac{\partial^2 P_1}{\partial x^2}
\label{eq1.2} \\
G &=& \frac{k_BT}{m\gamma}.
\label{eq1.3}
\end{eqnarray}

\noindent The structure of the diffusion constant $G$ is easily explained from the Langevin equations with heavy damping (i.e., $\gamma$ large) which can be written as

\begin{eqnarray}
\dot{x} &=& v \nonumber \\
\dot{v} &=& -\gamma v + \xi(t) \nonumber \\
\Rightarrow \dot{x} &=& \frac{1}{\gamma} \xi(t)
\label{eq1.4}
\end{eqnarray}

\noindent where, in the last step we have dropped the inertial term (i.e., $\ddot{x}$), as a consequence of heavy damping. Here $\xi(t)$ is the fluctuating force coming from the thermal bath, and is identified by the average $\langle \xi(t) \rangle = 0$ and the two-time Markovian correlation $\langle \xi(t) \xi(t') \rangle = \frac{2k_BT \gamma}{m}\delta(t - t')$. The last line of Eq.(\ref{eq1.4}) directly leads to the diffusion equation (\ref{eq1.3}). The standard process of solving the diffusion equation with given initial distribution is through the method of Fourier transformation. Introducing the Fourier transfrom of $P_1(x,t)$ as $\tilde{P}_1(k,t)$ we write

\begin{equation}
P_1 (x,t) = \frac{1}{\sqrt{2\pi}} \int_{-\infty}^{\infty} dk \tilde{P}_1 (k,t) e^{ikx}
\label{eq1.5}
\end{equation}

\noindent which when placed back in Eq.(\ref{eq1.2}) leads to the solution

\begin{equation}
\tilde{P}_1 (k,t) = \tilde{P}_1 (k,0) e^{-Gk^2t}
\label{eq1.6}
\end{equation}

\noindent where

\begin{equation}
\tilde{P}_1 (k,0) = \frac{1}{\sqrt{2\pi}} \exp \left( -\frac{1}{2} \sigma^2 k^2 \right).
\label{eq1.7}
\end{equation}

\noindent Incorporating Eqs.(\ref{eq1.6}) and (\ref{eq1.7}) in Eq.(\ref{eq1.5}) we finally get the distribution function for the first phase of the motion as

\begin{eqnarray}
P_1(x,t)  = \frac{1}{\sqrt{\pi(2\sigma^2 + 4Gt)}} \exp\left( -\frac{x^2}{2\sigma^2 + 4Gt} \right).
\label{eq1.8}
\end{eqnarray}

\noindent
It is clear from the structures of eqs.(\ref{eq1.1}) and (\ref{eq1.8}) that, with time the distribution flattens out, as a result of pure diffusion.

\noindent \underline{{\it The 2nd Phase of the Dynamics}:} After allowing the particles to spread for a certain length of time, say $t=\tau$, we now suddenly set up a harmonic trap about the origin. This triggers the second phase of the dynamics, which is hence governed by a different probability function $P_2(x,t)$ that obeys the Fokker-Planck equation

\begin{equation}
\frac{\partial P_2}{\partial t'} = \frac{1}{m\gamma} \frac{\partial}{\partial x} \{ V'(x) P_2 \} + G\frac{\partial^2 P_2}{\partial x^2}
\label{eq1.9}
\end{equation}

\noindent where, we have used a new time variable $t'$, so that $t = \tau + t'$. For solving this equation, we need an initial condition. The initial time instant for this second phase of motion is $t=\tau$. The value of $P_1(x,t)$ at $t=\tau$ provides that requisite initial distribution for this phase of the motion. Thus,

\begin{equation}
P_1(x,\tau) = P_2(x,0).
\label{eq1.10}
\end{equation}

\noindent Invoking, again, the Fourier transform as

\begin{equation}
P_2 (x,t) = \frac{1}{\sqrt{2\pi}} \int_{-\infty}^{\infty} dk \tilde{P}_2 (k,t) e^{ikx}
\label{eq1.11}
\end{equation}

\noindent and using it in Eq.(\ref{eq1.9}) we obtain

\begin{equation}
\tilde{P}_2 (k,t) = \tilde{P}_2 (k,0) \exp \left[ -Gk^2t' + \frac{iV'(x) t'}{m\gamma}k + \frac{V''(x) t'}{m\gamma} \right]
\label{eq1.12}
\end{equation}

\noindent with

\begin{equation}
\tilde{P}_2 (k,0) = \frac{1}{\sqrt{2\pi}} \exp \left[ -\frac{1}{2} \left( \sigma^2 k^2 + 2G\tau \right)\right].
\label{eq1.13}
\end{equation}

\noindent
Combining Eqs.(\ref{eq1.11})-(\ref{eq1.13}) we get the final expression for the probability distribution in the second stage of the dynamics as

\begin{eqnarray}
P_2(x,t') &=&  \frac{e^{V''(x)t' / m\gamma}}{\sqrt{\pi(2\sigma^2 + 4G(\tau+t'))}} \nonumber \\
&\times& \exp \left[ - \frac{\left( x + \frac{V'(x) t'}{m\gamma} \right)^2}{2\sigma^2 + 4G(\tau+t')} \right].
\nonumber \\
\label{eq1.14}
\end{eqnarray}

\noindent \underline{{\it Survival Probability and First Passage Distribution}:} With all cards on the table, we now ask the question: what are the survival probability and the first passage distribution for particles from within a prescribed region (say, from $-a <x < a$) to return to the origin (which is the centre of the harmonic trap suddenly established at time $t = \tau$)? In this work, we are not considering the existence of any absorbing barrier anywhere; it is just the time of first reaching the origin that concerns us here. The standard prescription of finding the first-passage (or first-reaching) distribution is to evaluate the survival probability first, followed by taking the negative of the time-derivative of the survival probability. The survival probability, $S(t')$, for particles confined within the region $-a < x < a$ is given by

\begin{equation}
S(t') = \int_{-a}^a dx P_2(x,t')
\label{eq1.15}
\end{equation}

\noindent and the first passage distribution $T(t')$ is given by

\begin{equation}
T(t') = -\frac{\partial S(t')}{\partial t'}.
\label{eq1.16}
\end{equation}

\noindent The integration for the survival probability has been done from $-a$ to $+a$ to imply that this is basically a sum of two different integrals: one from $-a \to 0$ and the other from $a \to 0$. Had we begun with an initial $\delta$-function distribution (as is commonly done for diffusion problems) about some point, say $x = x_0 > 0$, then the integral for survival probability (and hence first passage distribution) between the limits $-a \to 0$ would have been redundant, because the particles could go to the negative side of the $x$-axis only after crossing the origin at least once. But since we are beginning with an initial symmetric (Gaussian) distribution in space, the integral for survival probability has to cater to both sides of the origin. The expressions are formidably large and hence have been listed in the Appendix. We give the plots in Figs.1 to 4. In Fig.1 the normalized survival probability distribution $S(t')$ has been plotted against time $t'$ for three different values of the resetting time $\tau$, while in Fig.2 the same has been plotted for three different values of the initial half-width of space distribution $\sigma$. In Figs. 3 and 4, the plots of the first passage distribution have been given for different values of $\tau$ and $\sigma$ respectively. In all these figures, the mass ($m$) and the frequency ($\omega$) have been kept at unity, while the (heavy) damping constant has been chosen as $\gamma = 50$, so that the heavy-damping approximation (i.e., the time-scale defined by the reciprocal of $\gamma$ is much smaller relative to the other time scales, viz., $\tau$ and $t'$) works reliably well upto a time duration $t = \tau + t' = 5 + 10$, as is seen in the plots of Figs. 1 to 4. The integration limits of Eq.(\ref{eq1.15}) have been fixed at $a = \pm 1$.

\begin{figure}[h!]
\includegraphics[width=0.450\textwidth,angle=0,clip]{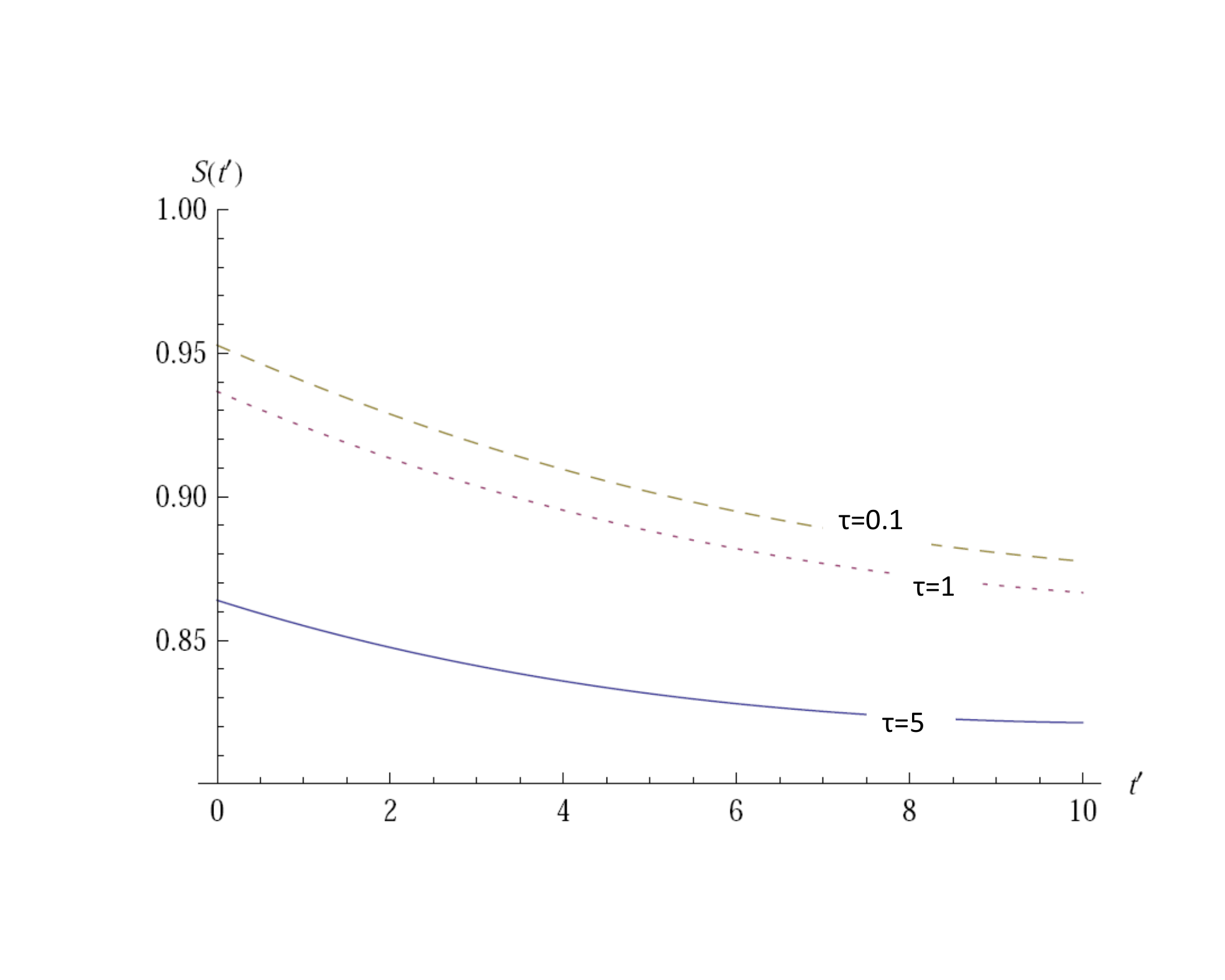}
\caption[]{Survival probability for three different values of $\tau$. Here $\sigma = 0.5$. }
\end{figure}

\begin{figure}[h!]
\includegraphics[width=0.450\textwidth,angle=0,clip]{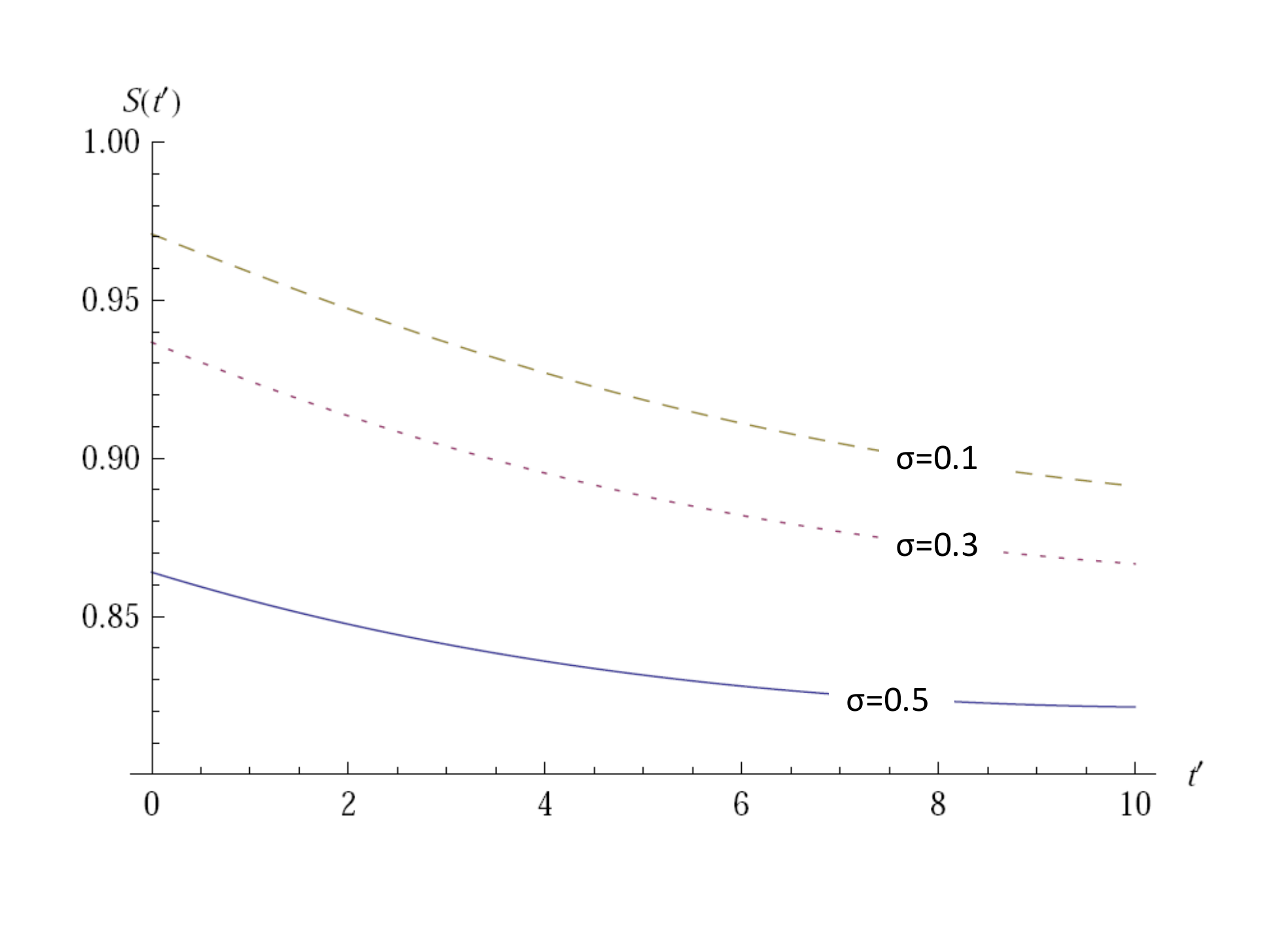}
\caption[]{Survival probability for three different values of $\sigma$. Here $\tau = 5$.}
\end{figure}

\begin{figure}[h!]
\includegraphics[width=0.450\textwidth,angle=0,clip]{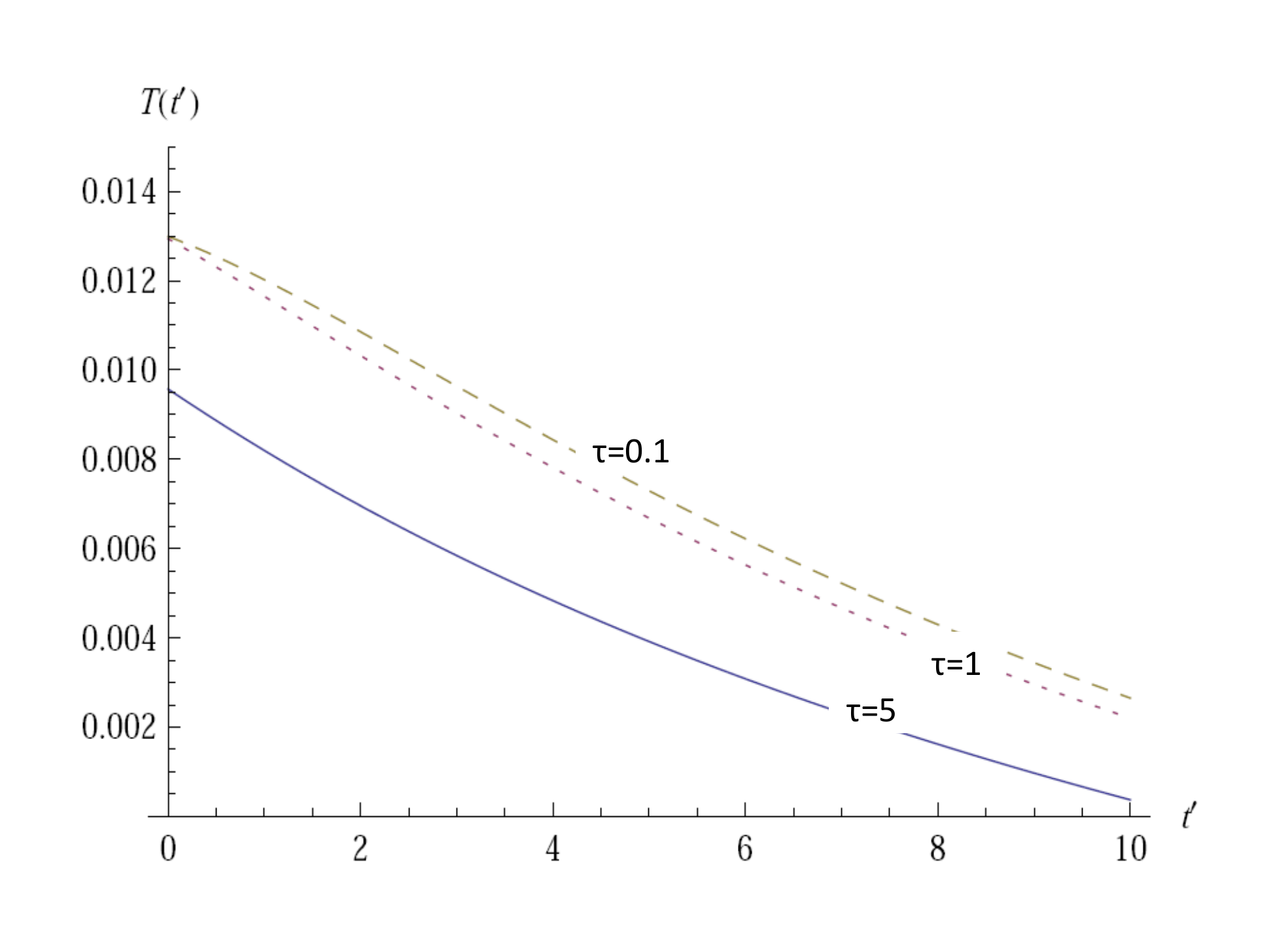}
\caption[]{First passage distribution for three different values of $\tau$ with $\sigma = 0.5$}
\end{figure}

\begin{figure}[h!]
\includegraphics[width=0.450\textwidth,angle=0,clip]{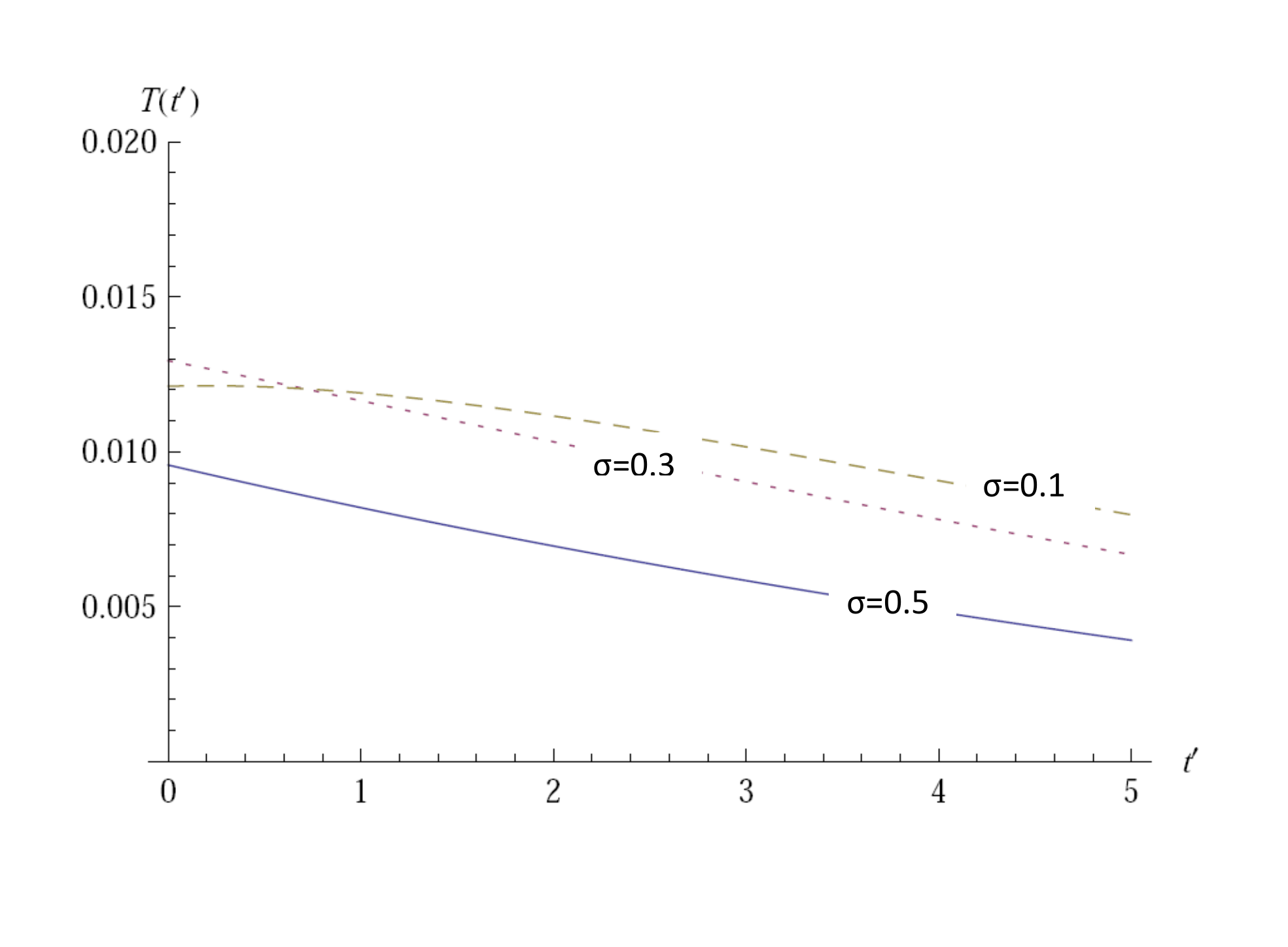}
\caption[]{First passage distribution for three different values of $\sigma$ with $\tau = 5$.}
\end{figure}

\noindent We note that a greater value of $\tau$ or a greater value of $\sigma$ are in a sense synonymous, because both imply a wider distribution at the beginning of the second phase of the dynamics: the former means that the distribution has been allowed to spread more by delaying the resetting order, and the latter implies that we have started out with a wider distribution. In both cases the survival probability decreases with increasing values of these parameters. This means that within a given domain, say $-1 \leq a \leq 1$, more spreading of the distribution takes the particles outside this domain thus reducing the survival probability within this domain.

\renewcommand{\theequation}{2.\arabic{equation}}
\setcounter{equation}{0}
\section{II. DETERMINISTIC RETURN: A KRAMERS-LIOUVILLE PROCESS}

\noindent We now formulate the same problem in a more complicated set-up where the velocity gets involved as a result of low damping. In the first phase of the dynamics, there is a free stochastic evolution governed by the Kramers equation. On the contrary, in the second phase, there is no thermal bath and the motion under the harmonic potential is purely deterministic and therefore governed by Liouville equation, as explained in the introduction. It should be clear that, the deterministic phase of the process has a randomness (both in position as well as velocity) built into it right at the point of its commencement, which is decided by the resetting time. If the resetting order is given at time $t= \tau$, then the Langevin equations for this combination of processes can be written as

\begin{eqnarray}
\dot{x} &=& v \label{eq2.1} \\
\dot{v} &=& \left[ -\frac{V'(x)}{m} \right]. \Theta (t - \tau) + [\xi(t) - \gamma v]. \Theta (\tau - t)
\label{eq2.2}
\end{eqnarray}

\noindent where, $m$ is the mass of the particle, $V(x)$ is the resetting potential (which we shall take to be harmonic in course of our calculations) and $\gamma$ is the damping (from a thermal bath) that the particle suffers during the stochastic phase of its motion under the action of the Gaussian white noise $\xi(t)$ characterized by its mean as $\langle \xi(t) \rangle = 0$ and by its two-time correlation function as $\langle \xi(t) \xi(t') \rangle = \frac{2\gamma k_B T}{m} \delta (t - t')$, where $T$ denotes the temperature of the thermal bath through which the Brownian particle moves during the stochastic phase of its motion, $k_B$ being the Boltzmann constant. The $\Theta$-functions (or heaviside functions) carry their usual meanings, viz., $\Theta(z) = 1$ for $z \geq 0$, and $\Theta(z) = 0$ otherwise. The appearances of these functions make it clear that the dynamics is governed by the potential term only after the resetting order is given at $t = \tau$. A probabilistic description of this process can also be formulated by writing down a Fokker-Planck equation from the above Langevin equations. In the first term on the right hand side of Eq.(\ref{eq2.2}), writing $\Theta (t - \tau) = 1 - \Theta (\tau - t)$, the equation for the evolution of probability function $P \equiv P(x,v,t)$ in phase space can be written down directly as

\begin{eqnarray}
\frac{\partial P}{\partial t} &=& - \frac{\partial}{\partial x} (vP) - \frac{\partial}{\partial v} \left[ \left\{
-\frac{V'(x)}{m} + \frac{V'(x)}{m}.\Theta(\tau - t) \right. \right. \nonumber \\
&-& \left. \left. \gamma v \Theta(\tau - t) \right\} P\right] + \Theta(\tau - t) \frac{\gamma k_BT}{m} \frac{\partial^2 P}{\partial v^2}.
\label{eq2.3}
\end{eqnarray}

\noindent This is a Kramers equation that can be split up into two separate equations valid for the two different temporal regimes. For $t < \tau$, we have the Krammers' equation for a free Brownian particle as

\begin{eqnarray}
\frac{\partial P_s}{\partial t} &=& -\frac{\partial}{\partial x}(vP_s) + \gamma \frac{\partial}{\partial v}(vP_s) + \frac{\gamma k_B T}{m} \frac{\partial^2 P_s}{\partial v^2} \label{eq2.4}
\end{eqnarray}

\noindent and for $t > \tau$, we have a pure Liouville equation for the probabilistic evolution of a particle under the action of a potential $V(x)$ as

\begin{equation}
\frac{\partial P_d}{\partial t} = -v \frac{\partial P_d}{\partial x} + \frac{V'(x)}{m} \frac{\partial P_d}{\partial v}.
\label{eq2.5}
\end{equation}

\noindent
The subscripts ``s" and ``d" denote the terms ``stochastic" and ``deterministic" respectively.

\noindent
We again do our calculations by assigning an initial probability distribution in phase space, which is Gaussian in both space and velocity coordinates. Thus,

\begin{equation}
P_s(x,v,0) = \frac{1}{2\pi \sigma} \sqrt{\frac{m}{k_BT}} \exp \left( -\frac{mv^2}{2k_BT} \right) \exp \left( -\frac{x^2}{2\sigma^2} \right)
\label{eq2.6}
\end{equation}

\noindent
where, the velocity part is a Maxwellian distribution, and the spatial part, as before, has a half-width $\sigma$, while the overall constant factor sitting in front ensures that the full phase-space distribution is normalized at $t=0$. Introducing, this time, a two-variable Fourier transform of the probability distribution as

\begin{equation}
P_s(x,v,t) = \left( \frac{1}{\sqrt{2\pi}} \right)^2 \int_{-\infty}^{\infty} \int_{-\infty}^{\infty} dx dv \tilde{P}_s(k,u,t) e^{i(kx + uv)}
\label{eq2.7}
\end{equation}

\noindent
in Eq.(\ref{eq2.4}) above, we have the differential equation in fourier space as

\begin{equation}
\frac{\partial \tilde{P}_s}{\partial t} = (-ikv + i\gamma uv + \gamma - \Gamma u^2) \tilde{P}_s
\label{eq2.8}
\end{equation}

\noindent where $\tilde{P}_s$ has to be understood as $\tilde{P}_s (k,u,t)$ (as opposed to $P_s \equiv P_s(x,v,t)$) and the constant $\Gamma = \gamma k_BT / m$. The solution of Eq.(\ref{eq2.8}) is easily obtained as

\begin{equation}
\tilde{P}_s(k,u,t) = \tilde{P}_s(k,u,0)e^{(-ikv + i\gamma uv + \gamma - \Gamma u^2)t}
\label{eq2.9}
\end{equation}

\noindent where

\begin{equation}
\tilde{P}_s(k,u,0) = \frac{1}{2\pi} \exp\left( -\frac{\sigma^2 k^2}{2} - \frac{\Gamma}{2\gamma}u^2 \right).
\label{eq2.10}
\end{equation}

\noindent
Putting Eqs.(\ref{eq2.9}) and (\ref{eq2.10}) back in Eq.(\ref{eq2.7}), we obtain, after some straight-forward algebra involving Gaussian integrations,

\begin{eqnarray}
P_s (x,v,t) &=& \frac{1}{2\sqrt{2}\pi\sigma} \frac{e^{\gamma t}}{\sqrt{\Gamma(t + \frac{1}{2\gamma})}} \nonumber \\
&\times& \exp \left[ -\frac{(x - vt)^2}{2\sigma^2} - \frac{(1 + \gamma t)v^2}{4\Gamma(t + \frac{1}{2\gamma})} \right].
\label{eq2.11}
\end{eqnarray}

\noindent
We now make the weak-damping approximation. This means, that for any time $t \leq \tau$ (where $\tau$ is the time instant when the resetting order is suddenly given), we have $\gamma t << 1$. For the stochastic part of the motion we have $t < \tau$, and hence, in Eq.(\ref{eq2.11}) we shall retain only terms linear in $\gamma t$. Thus, expanding the terms like $(t + 1/2\gamma)^{-1/2}$, $(t + 1/2\gamma)^{-1}$ and $e^{\gamma t}$ in powers of $\gamma t$ and retaining only the linear terms we have

\begin{eqnarray}
P_s(x,v,t) = \frac{1}{2\pi\sigma} \sqrt{\frac{\gamma}{\Gamma}} \exp \left[ -\frac{(x - vt)^2}{2\sigma^2} - \frac{\gamma}{2\Gamma}v^2 \right].
\label{eq2.12}
\end{eqnarray}

\noindent It is interesting to note from Eqs.(\ref{eq2.6}) and (\ref{eq2.12}) that, in this weak damping regime, the exponential part of the phase-space probability distribution retains its Gaussian structure, with the variable $x$ replaced by $(x - vt)$. The long time dynamics is dictated by the $-t^2$ term sitting in the exponential, thus ensuring that this probability function never diverges.

\noindent \underline{{\it The Second Phase with Deterministic Return}:}  We now move to the second part of the motion. The governing equation for this part is Eq.(\ref{eq2.5}) with the initial condition provided by the condition

\begin{equation}
P_d(x,v,0) = P_s(x,v,\tau)
\label{eq2.13}
\end{equation}

\noindent where, for this particular process, in the second phase of the motion there are no damping and noise terms to complement each other. A harmonic potential $V(x) = \frac{1}{2} m \omega^2 x^2$ is suddenly set up at $t=\tau$, so that for $t> \tau$, the probability evolution equation is the Liouville equation, given in Eq.(\ref{eq2.5}), viz.,

\begin{equation}
\frac{\partial P_d}{\partial t'} = -v \frac{\partial P_d}{\partial x} + \frac{V'(x)}{m} \frac{\partial P_d}{\partial v}.
\label{eq2.14}
\end{equation}

\noindent written in terms of the new time variable $t'$ such that $t= \tau$ corresponds to the instant $t' = 0$. Again, introducing the Fourier transform $\tilde{P}_d (k,u,t')$ as

\begin{equation}
P_d(x,v,t') = \frac{1}{2\pi} \int_{-\infty}^{\infty} \int_{-\infty}^{\infty} dk du \tilde{P}_d(k,u,t') e^{i(kx + uv)}
\label{eq2.15}
\end{equation}

\noindent we have,

\begin{equation}
\tilde{P}_d(k,u,t') = \tilde{P}_d(k,u,0) \exp \left[ \left( -ikv + i \frac{V'(x)}{m}u \right)t' \right].
\label{eq2.16}
\end{equation}

\noindent Since Eq.(\ref{eq2.13}) holds for the respective Fourier transforms as well, i.e.,

\begin{equation}
\tilde{P}_s (k,u,\tau) = \tilde{P}_d(k,u,0)
\label{eq2.17}
\end{equation}

\noindent we have from Eq.(\ref{eq2.12}),

\begin{eqnarray}
\tilde{P}_s (k,u,\tau) &=& \frac{1}{4\pi^2 \sigma} \sqrt{\frac{\gamma}{\Gamma}} \int_{-\infty}^\infty \int_{-\infty}^\infty dx dv \nonumber \\
&\times& \exp \left[ -\left(\frac{x^2}{2\sigma^2} - \frac{xv\tau}{\sigma^2} + ikx\right) \right. \nonumber \\
&-& \left. \left(\frac{\tau^2}{2\sigma^2}v^2 + \frac{\gamma}{2\Gamma}v^2 + iuv \right) \right]
\label{eq2.18}
\end{eqnarray}

\noindent where, the terms in the exponential are grouped in this order, to facilitate the integration. Doing the $x$-integration first, with the terms in the first pair of parentheses and treating $v$ as a constant for this integration, and then doing the $v$-integration we get

\begin{eqnarray}
\tilde{P}_s (k,u,\tau) &=& \frac{1}{2\pi} \exp \left[ -\frac{\Gamma}{2\gamma} (u + k\tau)^2 - \frac{\sigma^2 k^2}{2} \right] \nonumber \\
&=& \tilde{P}_d (k,u,0).
\label{eq2.19}
\end{eqnarray}

\noindent Combining Eqs.(\ref{eq2.16})-(\ref{eq2.19}) with Eq.(\ref{eq2.15}) we get

\begin{eqnarray}
P_d (x,v,t') &=& \frac{1}{4\pi^2} \int_{-\infty}^\infty \int_{-\infty}^\infty dk du \nonumber \\
&\times& \exp \left[ -\frac{\Gamma}{2\gamma}u^2 - \frac{\Gamma k\tau}{\gamma}u + \frac{iV'(x)t'}{m}u + ivu \right. \nonumber \\
&-& \left. \frac{\sigma^2}{2}k^2 - \frac{\Gamma \tau^2}{2\gamma}k^2 - ivt'k + ixk \right].
\label{eq2.20}
\end{eqnarray}

\noindent
Doing the $u$-integration first with the terms in the second line of the above equation, followed by the $k$-integration with the remaining terms, we are led to the final expression for the probability distribution for this second phase of the dynamics as

\begin{eqnarray}
P_d (x,v,t') &=& \frac{1}{2\pi\sigma} \sqrt{\frac{\gamma}{\Gamma}} \exp \left[ -\frac{1}{2\sigma^2} \left\{ x - v(\tau + t') \right. \right. \nonumber \\
&-& \left. \left. \frac{V'(x)}{m} \tau t' \right\}^2 - \frac{\gamma}{2\Gamma} \left\{ v + \frac{V'(x)}{m}t' \right\}^2 \right]. \nonumber \\
\label{eq2.21}
\end{eqnarray}

\noindent
A quick comparison of Eqs.(\ref{eq2.12}) and (\ref{eq2.21}) reveals that if the potential term $V'(x)$ is switched off, then the later reduces to the former with $t = \tau + t'$. The influence of the parameter $\tau$ on the final form of the distribution is clear from the product term $\frac{V'(x)}{m} \tau t'$: the distribution becomes more flattened out for a larger $\tau$, which means that a delay in giving the resetting order allows the particles to spread more, thus in turn affecting the survival probability and the first passage distribution, which we now calculate.

\noindent \underline{{\it Survival Probability and First Passage Distribution}:} We again ask the question: what is the first passage distribution for particles from within a prescribed region (say, from $-a <x < a$ and, this time, with all possible velocities) to return to the origin (which is the centre of the harmonic trap suddenly established at time $t = \tau$)? The survival probability $S(t')$ evaluates to

\begin{equation}
S(t') = \int_{-a}^{a} dx \int_{-\infty}^{\infty} dv P_d(x,v,t').
\label{eq2.22}
\end{equation}

\noindent
The $v$-integration is a simple Gaussian, followed by the $x$-integration which yields an error function. Differentiating that with respect to $t'$ the first passage distribution $T(t') = -\frac{\partial S(t')}{\partial t'}$ is obtained, the expression for which along with that of the survival probability have been listed in the Appendix.

\begin{figure}[h!]
\includegraphics[width=0.450\textwidth,angle=0,clip]{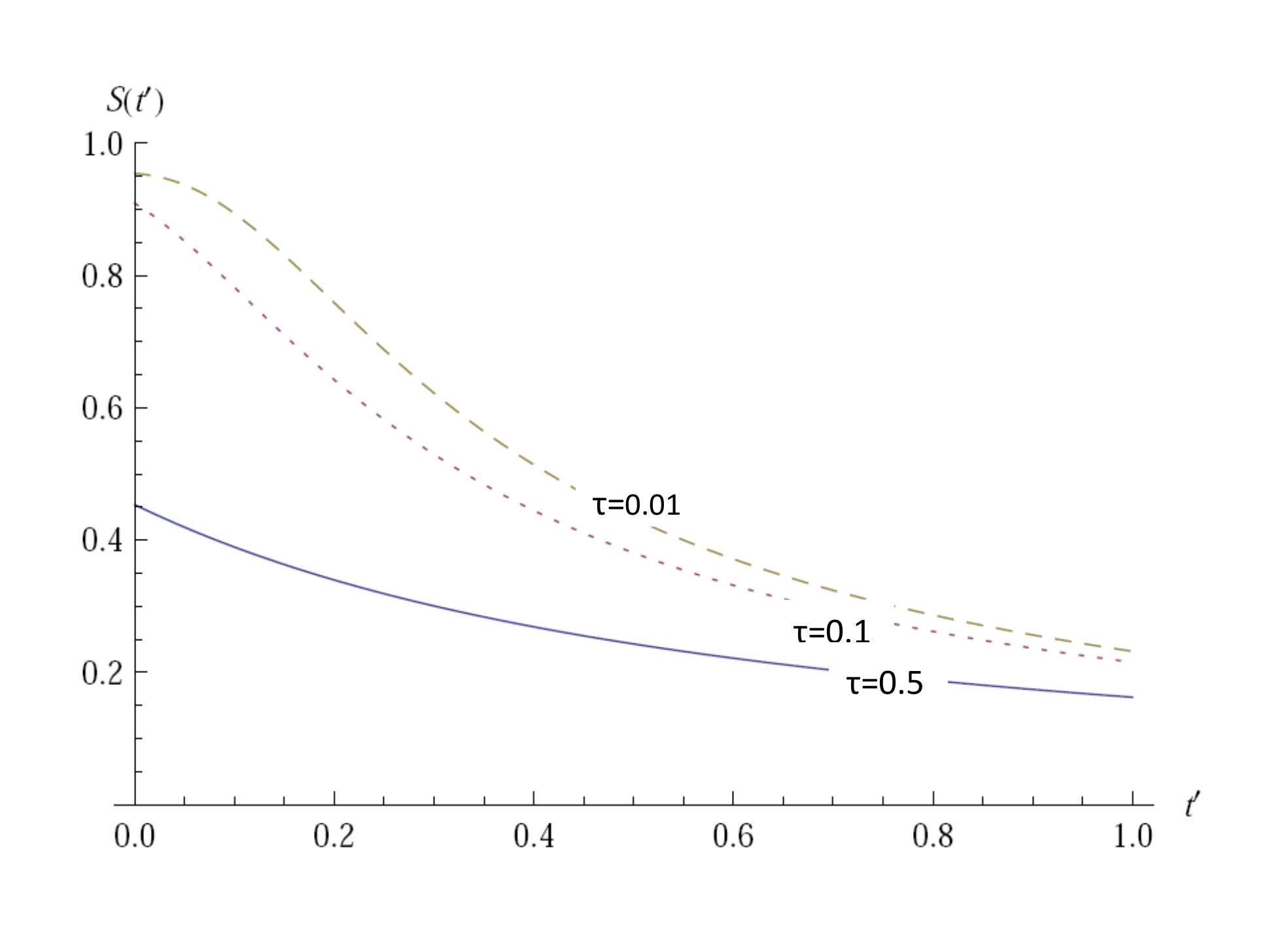}
\caption[]{Survival probability for three different values of $\tau$, with $\sigma = 0.5$ }
\end{figure}

\begin{figure}[h!]
\includegraphics[width=0.450\textwidth,angle=0,clip]{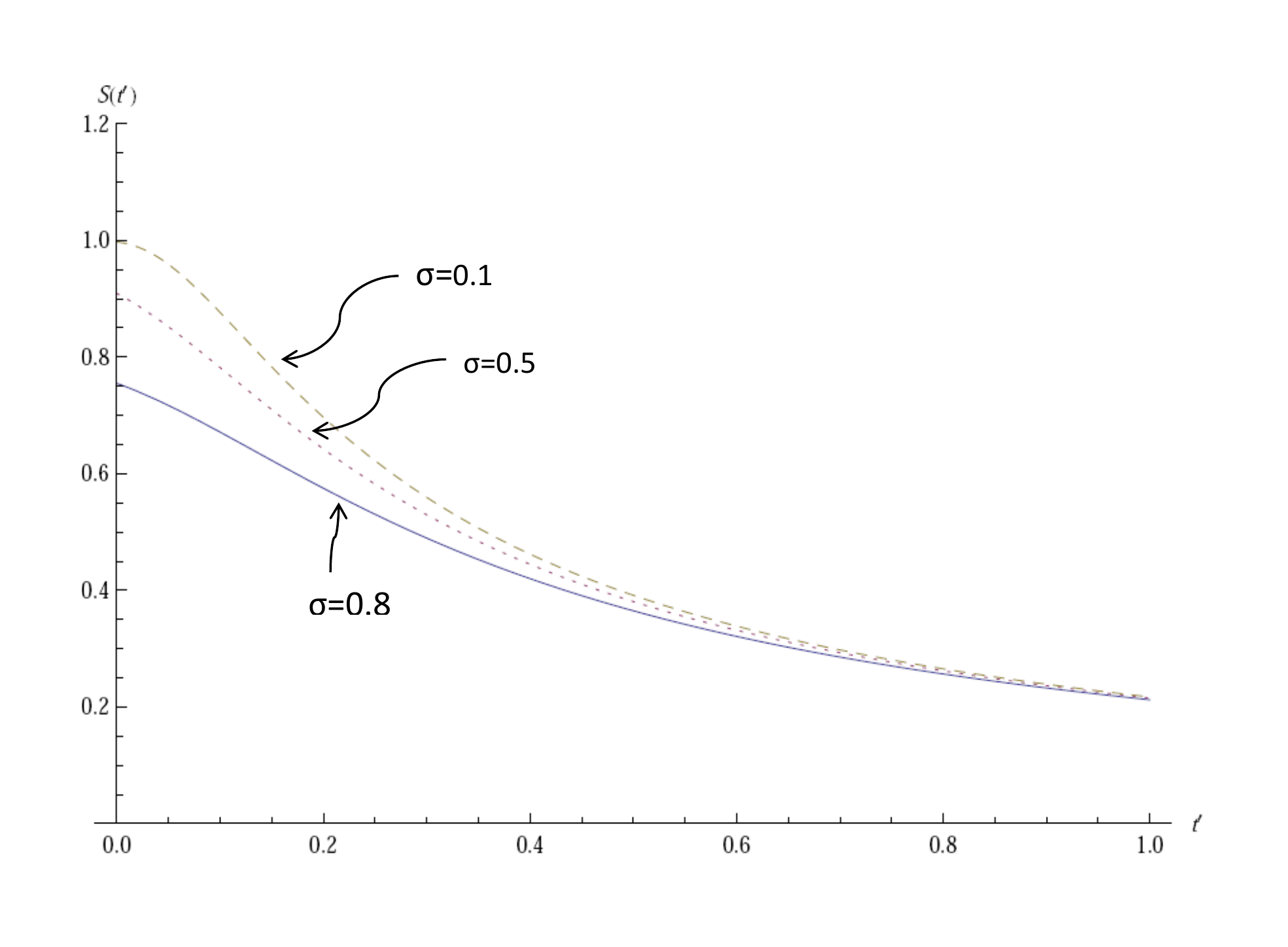}
\caption[]{Survival probability for three different values of $\sigma$ with $\tau = 0.1$. }
\end{figure}

\begin{figure}[h!]
\includegraphics[width=0.450\textwidth,angle=0,clip]{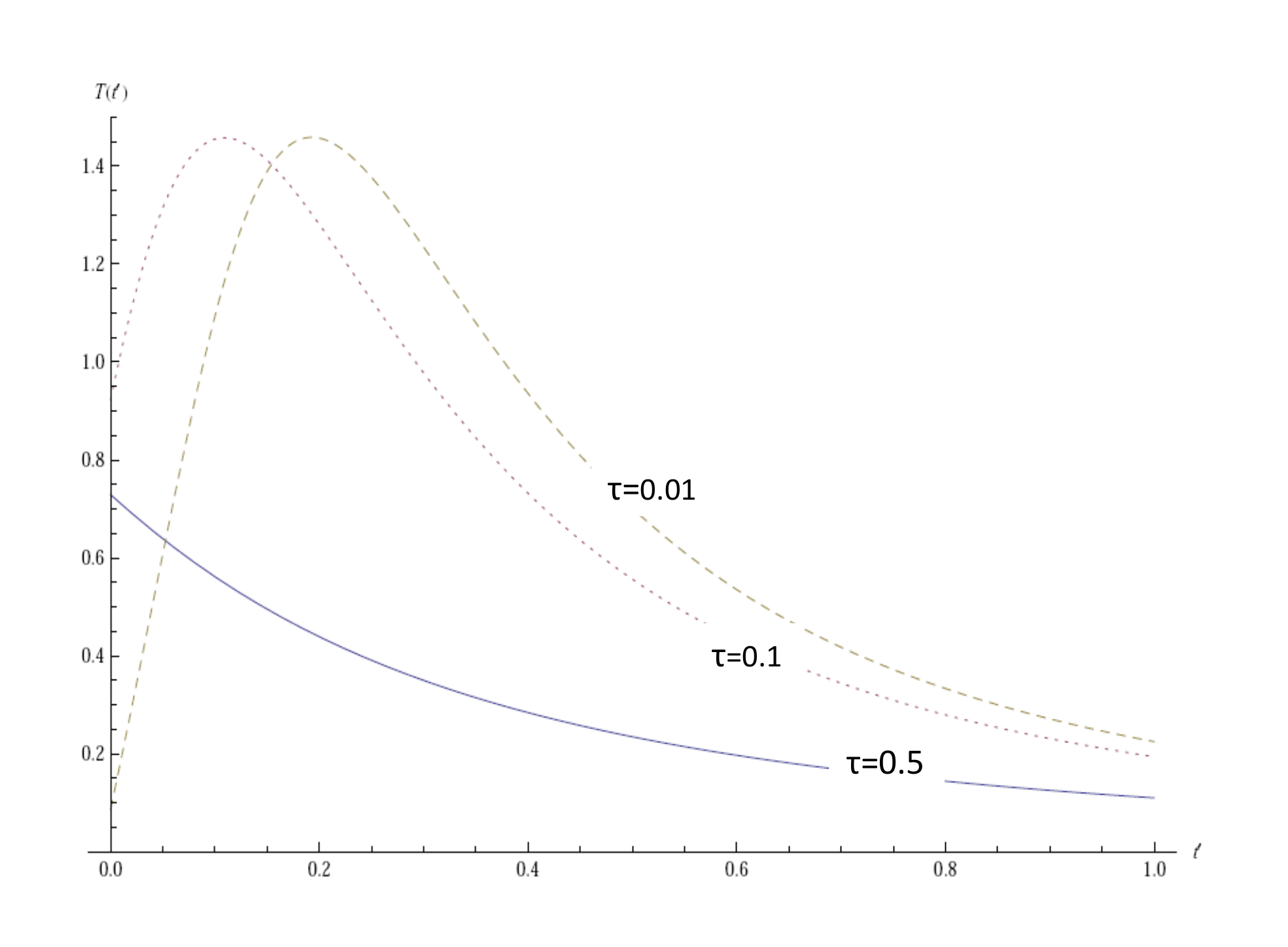}
\caption[]{First passage distribution for three different values of $\tau$ with $\sigma = 0.5$ }
\end{figure}

\begin{figure}[h!]
\includegraphics[width=0.450\textwidth,angle=0,clip]{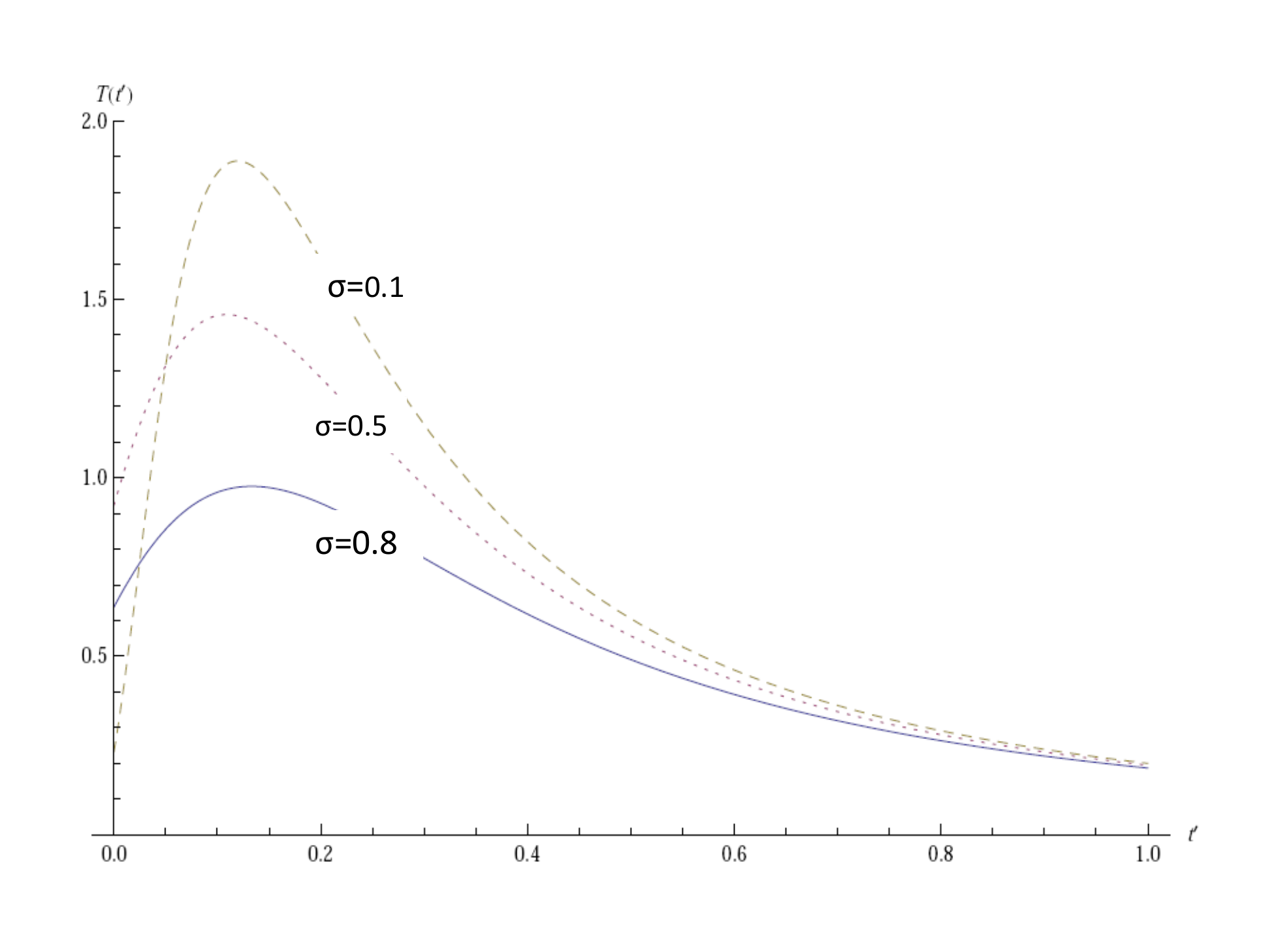}
\caption[]{First passage distribution for three different values of $\sigma$ with $\tau = 0.1$. }
\end{figure}

\noindent The plots of the normalized survival probability and the first passage distribution are given in Figs.5 to 8 for different values of $\tau$ and $\sigma$. We note that, for a fixed $t'$ the value of the function $S(t')$ is lesser for higher $\tau$. A higher value of $\tau$ means that in the stochastic phase of the motion, the distribution has been allowed to spread significantly from its initially assigned shape. Therefore, on giving the resetting order at a higher $\tau$, the possibility of return is much weakened due to greater spread. The decay of the first passage distribution with $t'$ is also similarly explained. As before, the constants $m$ and $\omega$ have been kept at unity. For weak damping, the damping constant has been chosen as $\gamma = 0.1$, such that the reciprocal of $\gamma$ describes a time-scale sufficiently large in comparison to the observational time-scale (which is roughly $t = \tau + t' = 0.5 + 1$), within which the plots give physically relevant results.

\renewcommand{\theequation}{3.\arabic{equation}}
\setcounter{equation}{0}
\section{III. The Low Damping Case: A pure Kramers process}

\noindent This process, as the name implies, takes place in a low damping environment throughout. The essential procedure remaining same, we state the main results in this Section. Equations for both the phases of the motion are governed by Kramers equation, the only difference being that, in the first phase, there is no potential while in the beginning of the second phase the harmonic potential is suddenly set-up. Operationally, the dynamics of the first phase is the same as the first phase of the Kramers-Liouville process described in the previous section. Thus the governing equation for the first phase is just Eq.(\ref{eq2.4}), the solution of which in the low damping approximation is given by Eq.(\ref{eq2.12}). For the second phase, the governing equation is

\begin{eqnarray}
\frac{\partial P_d}{\partial t} &=& -v\frac{\partial P_d}{\partial x} + \gamma \frac{\partial}{\partial v}(vP_d)
 \nonumber \\
 &+& \frac{V'(x)}{m}\frac{\partial P_d}{\partial v} + \Gamma \frac{\partial^2 P_d}{\partial v^2}
\label{eq3.1}
\end{eqnarray}

\noindent where, we have maintained the nomenclature of the previous section, although in this process both the phases are stochastic in nature. Applying the matching condition of Eq.(\ref{eq2.13}) and invoking the same Fourier transformation as given by Eq.(\ref{eq2.15}) we arrive at the integral

\begin{eqnarray}
P_d(x,v,t') &=& \frac{e^{\gamma t'}}{4\pi^2} \int_{-\infty}^\infty \int_{-\infty}^\infty dk du \nonumber \\
&\times& \exp \left[ -\frac{1}{2} \sigma^2 k^2 - \frac{\Gamma}{2\gamma} (u + k\tau)^2 - ikvt' \right.
\nonumber \\
&+& \left. i\gamma u vt' - \Gamma u^2 t' + ikx + iuv + \frac{iV'(x) t'}{m}u \right].
\nonumber \\
\label{eq3.2}
\end{eqnarray}

\begin{figure}[h!]
\includegraphics[width=0.450\textwidth,angle=0,clip]{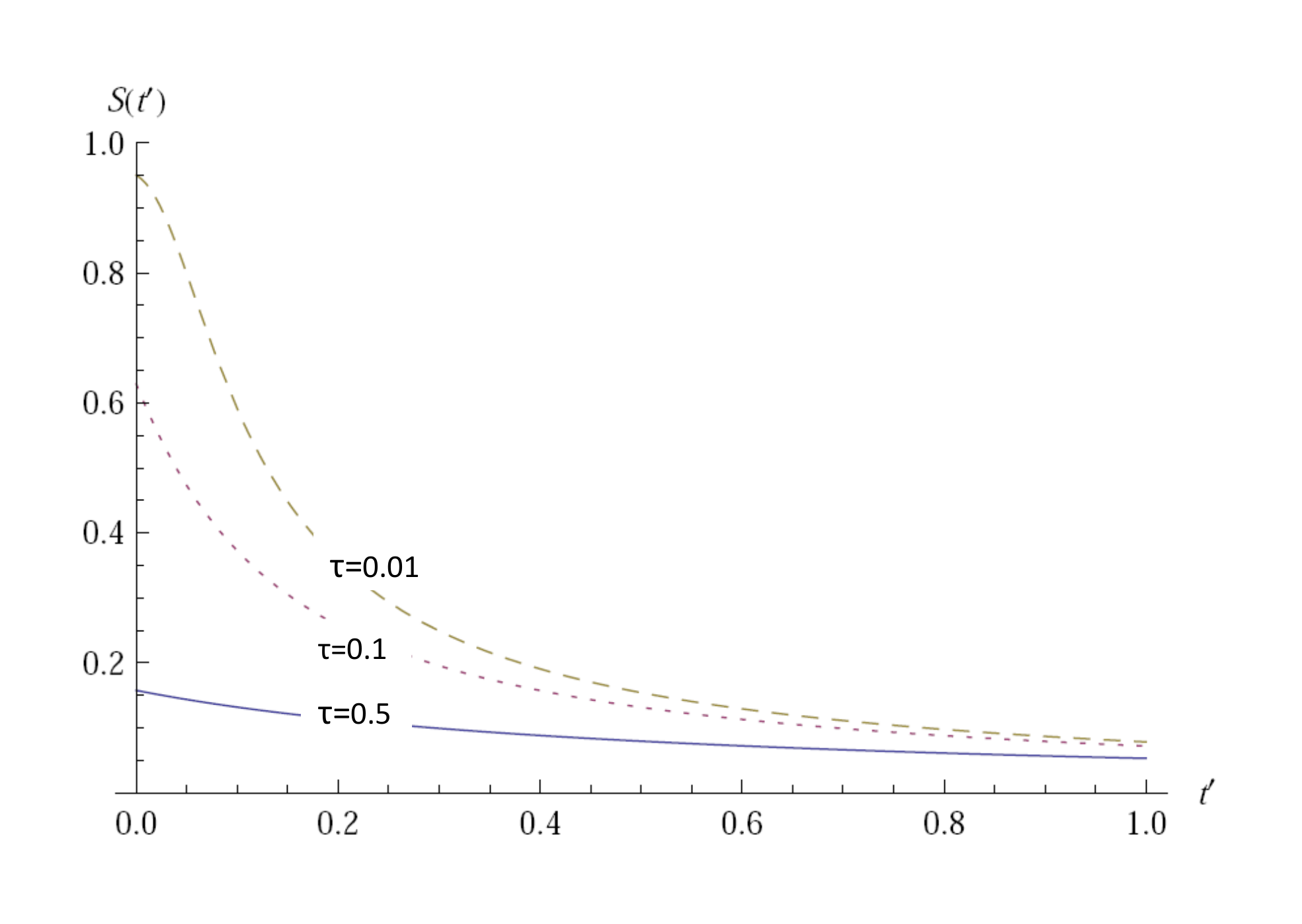}
\caption[]{Survival probability for three different values of $\tau$. }
\end{figure}

\begin{figure}[h!]
\includegraphics[width=0.450\textwidth,angle=0,clip]{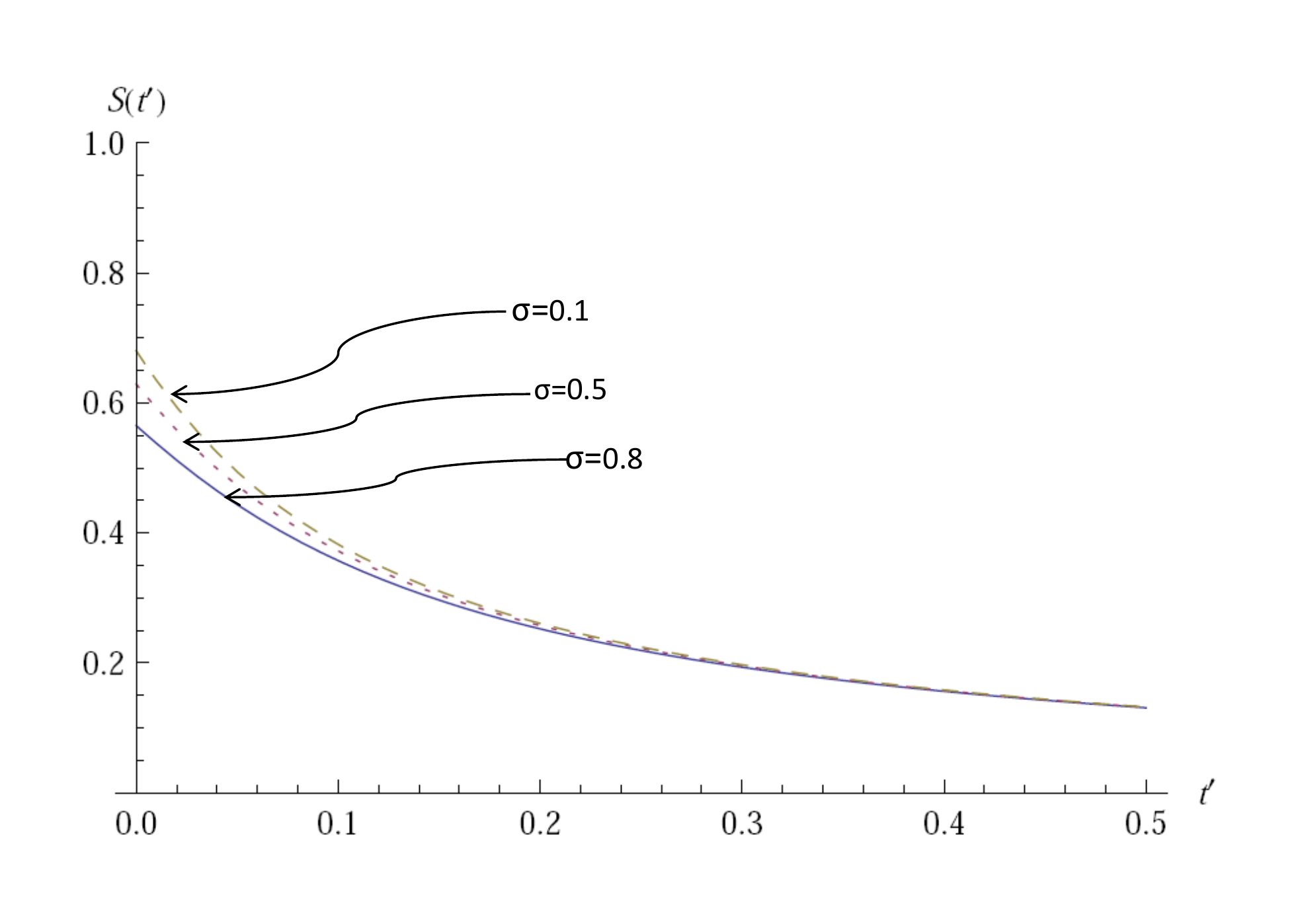}
\caption[]{Survival probability for three different values of $\sigma$. }
\end{figure}

\begin{figure}[h!]
\includegraphics[width=0.450\textwidth,angle=0,clip]{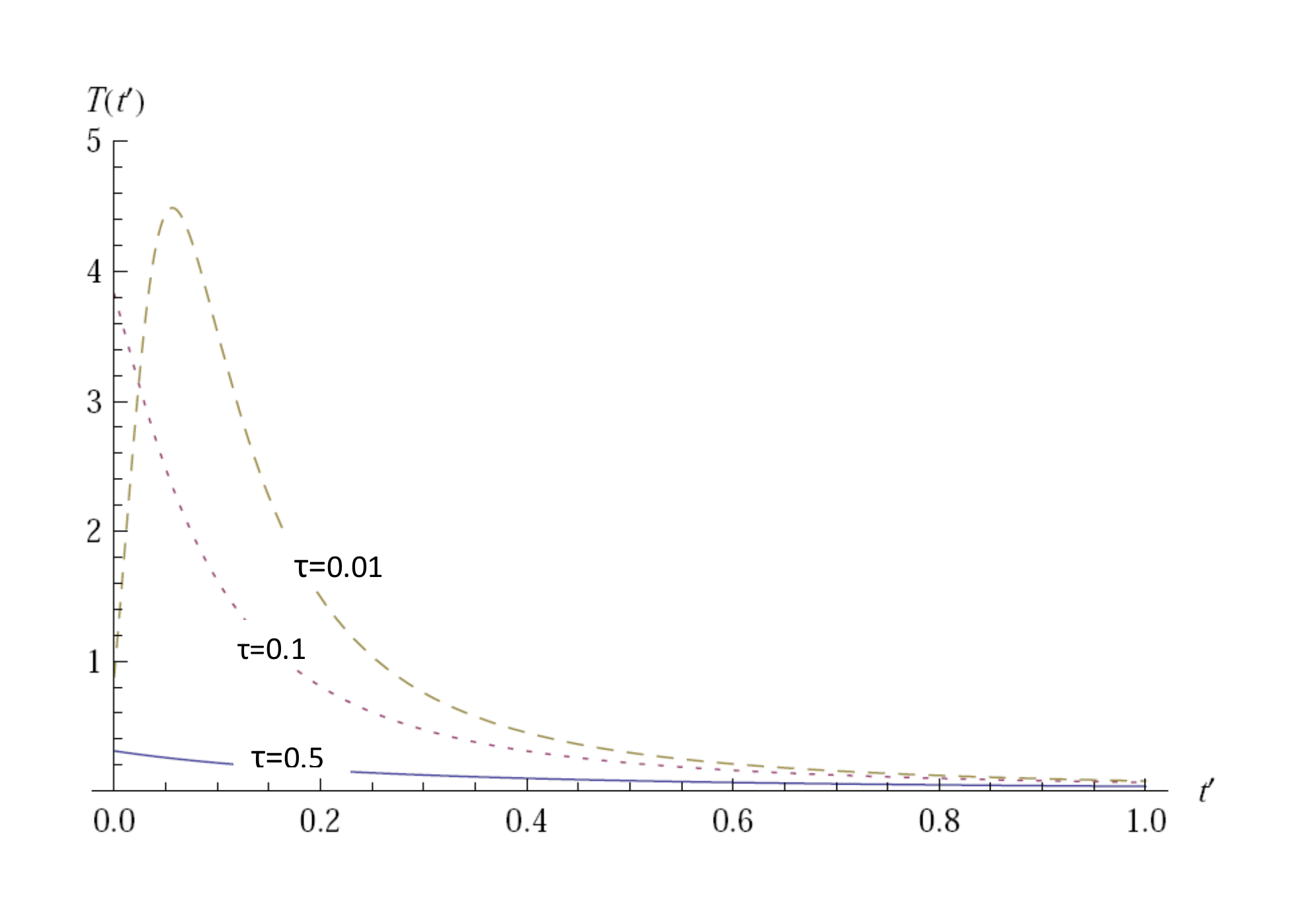}
\caption[]{First passage distribution for three different values of $\tau$. }
\end{figure}

\begin{figure}[h!]
\includegraphics[width=0.450\textwidth,angle=0,clip]{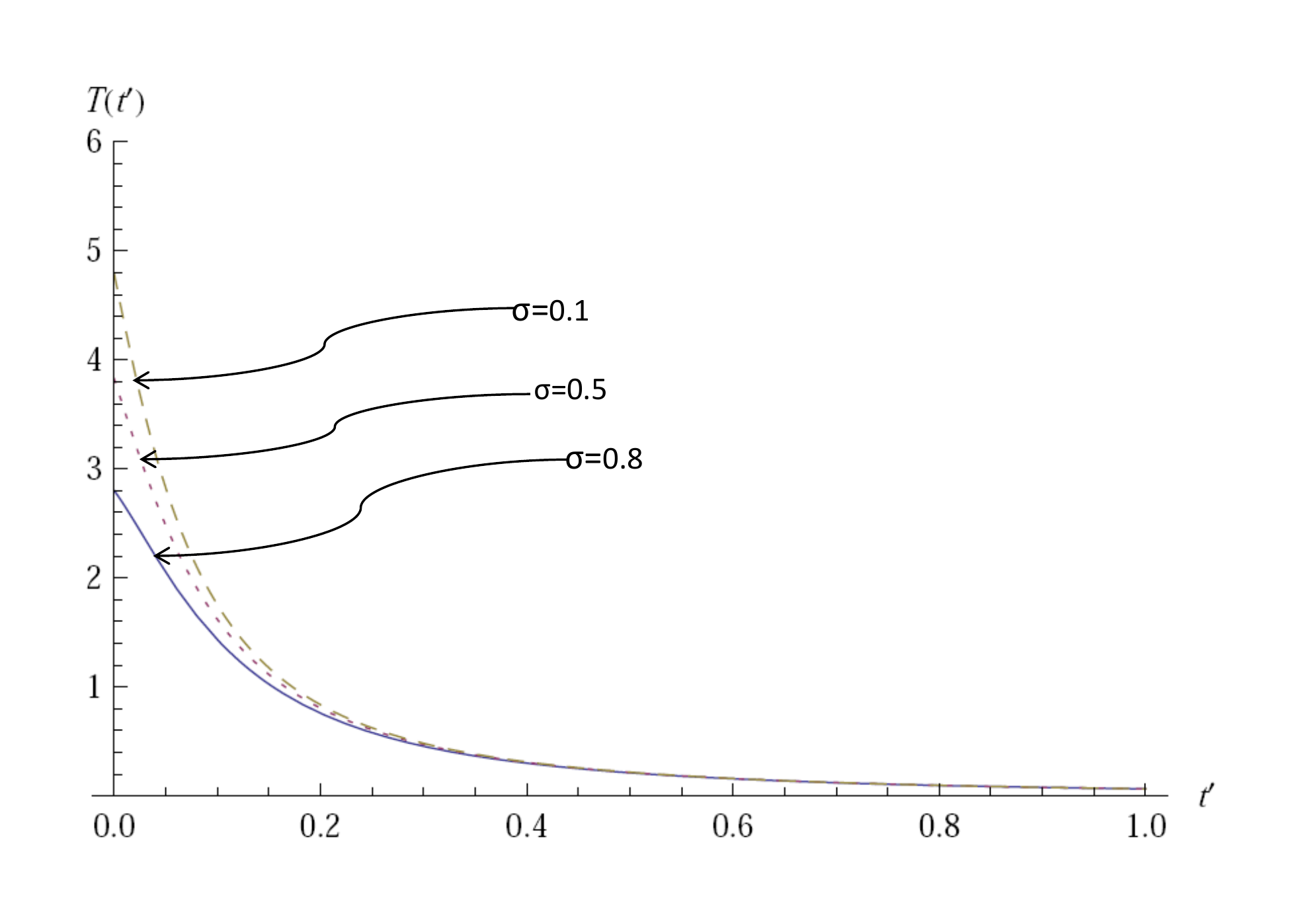}
\caption[]{First passage distribution for three different values of $\sigma$.  }
\end{figure}

\noindent
Performing the $u$-integration followed by the $k$-integration we finally arrive at the final expression for the phase space probability distribution $P_d(x,v,t')$ as

\begin{eqnarray}
P_d(x,v,t') &=& \frac{e^{\gamma t'}}{4\pi} \sqrt{\frac{1}{Z A(t')}}
\exp \left[ - \frac{B^2(t')}{A(t')}  \right.
\nonumber \\
&-& \left. \frac{1}{Z} \left\{ (x - vt') - \frac{2\Gamma \tau B(t')}{\gamma A(t')} \right\}^2 \right]
\label{eq3.3}
\end{eqnarray}

\noindent where,

\begin{eqnarray}
A(t') &=& 4\Gamma \left( t' + \frac{1}{2\gamma} \right)
\label{eq3.4} \\
B(t') &=& v(1 + \gamma t') + V'(x) t'
\label{eq3.5} \\
Z &=& 4 \left[ \frac{\sigma^2}{2} + \frac{\Gamma}{2\gamma} \left( 1 - \frac{2\Gamma}{\gamma A(t')} \right) \right].
\label{eq3.6}
\end{eqnarray}

\noindent
The plots of the normalized probability distribution and the first passage distribution are given for different values of $\tau$ and $\sigma$, as before [see Figs.9, 10, 11 and 12]. In this set, we have kept the value of the low damping coefficient at $\gamma = 0.01$. As we see, the basic patterns of variations for the graphs remain unaltered. As $\tau \to 0$, the survival probability approaches $1$, as is evident from the fact that if the distribution is not allowed to evolve at all, then one should find the particles within the prescribed domain with almost complete certainty. The hump seen in the behaviour of the first passage distribution is a characteristic of the low-damping scenario, as was noticed with the graphs of the Kramers-Liouville process also. With decreasing $\tau$ or decreasing $\sigma$, the duration and importance of the first phase of the dynamics shrinks to zero, vindicated by the fact that, the starting point of the first passage distribution lowers towards the origin, which is observed for standard one-step Brownian motion problems.

\noindent \underline{{\it Summary}:} In this paper we have studied the dynamics of stochastic processes that occur in two interconnected stages. The demarcation between the stages is characterized by a sudden setting up of a harmonic potential. Starting from a Gaussian probability distribution in phase space (or configuration space), particles start diffusing under the action of damping and stochastic forces till a certain time instant, called the resetting time, when a potential is set up about the origin to bring the particles back to the origin. We have calculated the survival probabilities and first passage distributions for particles from within a certain domain to return to the origin, under various damping environments. Familiar examples of such two-step processes are provided by recent studies on stochastic resetting problems, mostly aimed at studying the optimal returning rates of random searchers. Our angle of approach in this paper, however, is completely different and aims at studying the dynamics of both stages of the motion from an equation-of-motion point of view.

\renewcommand{\theequation}{A.\arabic{equation}}
\setcounter{equation}{0}
\section*{APPENDIX}
\noindent In this Appendix we enlist the expressions for the survival probabilities (given by $S(t')$) and first passage distributions (given by $T(t')$) for the three cases considered. In these expressions we have put $m=1$.

\subsection{For Heavy Damping: Pure Smoluchowski process}

\begin{eqnarray}
S(t') = \frac{\gamma e^{h(0,t')/\gamma}}{h(\gamma,t')}. \mathrm{Erf} \left[ \frac{h(\gamma,t')}{\gamma \sqrt{2f(t')}} \right]
\label{eqA.1}
\end{eqnarray}

\noindent and

\begin{eqnarray}
T(t') &=& \frac{2\gamma}{\sqrt{\pi} h(\gamma,t')} \exp \left[ \frac{h(0,t')}{\gamma} - \frac{h^2(\gamma,t')}{2a^2 f(t')} \right]
\nonumber \\
&\times& \left[ \frac{\omega^2}{\gamma \sqrt{2f(t')}} - \frac{G h(\gamma,t')}{\gamma \sqrt{2} [h(\gamma,t')]^{3/2}} \right]
\nonumber \\
&+& \frac{\omega^2 e^{h(0,t')/\gamma}}{h(\gamma,t')} \left( 1 - \frac{\gamma}{h(\gamma,t')} \right)
\nonumber \\
&\times& \mathrm{Erf} \left[ \frac{h(\gamma,t')}{\gamma \sqrt{2f(t')}} \right]
\label{eqA.2}
\end{eqnarray}

\noindent where

\begin{eqnarray}
f(t') &=& \sigma^2 + 2G(\tau + t'),
\nonumber \\
h(\gamma,t') &=& \gamma + \omega^2 t'.
\nonumber
\end{eqnarray}

\subsection{For Low Damping: Kramers-Liouville process}

\begin{eqnarray}
S(t') &=& \frac{1}{q(t')} \mathrm{Erf} \left[ \frac{\sqrt{\gamma} q(t')}{\sqrt{2p(t')}} \right]
\label{eqA.3}
\end{eqnarray}

\noindent and

\begin{eqnarray}
T(t') &=& \frac{2}{\sqrt{\pi} q(t')} \exp \left( -\frac{\gamma q^2(t')}{2p(t')}  \right)
\nonumber \\
&\times& \left[ \frac{\sqrt{2} \Gamma(\tau + t') q(t')}{\sqrt{2} [p(t')]^{3/2}}  -  \frac{\sqrt{2\gamma} \omega^2 t'}{\sqrt{p(t')}} \right]
\nonumber \\
&+& \frac{2\omega^2 t'}{[q(t')]^2} \mathrm{Erf} \left[ \frac{\sqrt{\gamma}q(t')}{\sqrt{2p(t')}} \right]
\label{eqA.4}
\end{eqnarray}

\noindent where

\begin{eqnarray}
p(t') &=& \gamma \sigma^2 + \Gamma (\tau + t')^2
\nonumber \\
q(t') &=& 1 + \omega^2 {t'}^2
\nonumber
\end{eqnarray}

\subsection{For Low Damping: Pure Kramers process}

\begin{eqnarray}
S(t') &=& \frac{e^{\gamma t'}}{w(t')} . \mathrm{Erf} \left[ \frac{\gamma w(t') \sqrt{y(t')}}{\sqrt{u(t')}} \right]
\label{eqA.5}
\end{eqnarray}

\noindent and

\begin{eqnarray}
T(t') &=& -\frac{2}{\sqrt{\pi w(t')}} \exp \left[ \gamma t' - \frac{\gamma^2 w^2(t') y(t')}{u(t')} \right]
\nonumber \\
&\times& \left[ \frac{\gamma(2v(t') - \gamma) \sqrt{y(t')}}{\sqrt{u(t')}} - \frac{s(t')}{2u^{3/2}(t')} \right]
\nonumber \\
&-& \left[ \frac{(2v(t') - \gamma) e^{\gamma t'}}{w^2(t')} - \frac{\gamma e^{\gamma t'}}{w(t')}\right]
\nonumber \\
&\times& \mathrm{Erf} \left[ \frac{\gamma w(t') \sqrt{y(t')}}{\sqrt{u(t')}} \right].
\label{eqA.6}
\end{eqnarray}

\noindent
where

\begin{eqnarray}
y(t') &=& 2 \Gamma \left( \frac{1}{\gamma} + 2t' \right)
\nonumber \\
z(t') &=& 4 \left[ \frac{\sigma^2}{2} + \left( \frac{\Gamma}{2\gamma}
- \frac{\Gamma^2}{\gamma^2}y(t') \right) \tau^2 \right]
\nonumber \\
u(t') &=& 2 \Gamma \tau {[(1 + \gamma t') + \gamma t' y ]}^2
\nonumber \\
&+& \gamma^2(1 + \gamma t')^2.y(t').z(t')
\nonumber \\
v(t') &=& \gamma + \omega^2 t'
\nonumber \\
w(t') &=& 1 + v(t').t'
\nonumber \\
s(t') &=& \gamma w(t') \sqrt{y(t')}.\left[ 2\gamma(2\Gamma \tau + y(t') ) \left\{ 2\Gamma \tau(1 + \gamma t')
\right. \right.
\nonumber \\
&+& \left. \left. \gamma y(t') t' \right\} + 2\gamma^3 (1 + \gamma t') y(t') z(t') \right].
\nonumber
\end{eqnarray}

\acknowledgements

\noindent \textbf{Acknowledgements}: Jyotipriya Roy and Debapriya Das are grateful to UGC (India) and Chitrak Bhadra is grateful to UGC-CSIR (India) for financial support. Jyotipriya, Chitrak, Debapriya and Dhruba are thankful to Professors Satya N. Majumdar, Abhishek Dhar and Sanjib Sabhapandit for allowing them to attend their excellent set of lectures on nonequilibrium statistical physics and stochastic processes in RRI, Bangalore in April 2014. Jyotipriya and Dhruba got acquainted with the subject of stochastic resetting from the authoritative set of lectures delivered by Professor Satya N. Majumdar in RRI, Bangalore during April 2013.

\end{document}